\documentclass{ptapap}
\usepackage{amsmath}
\usepackage{amssymb}
\usepackage{graphicx}
\usepackage{hyperref}
\usepackage[]{natbib}  
\usepackage{epstopdf}

\newcommand{\mass}{M$_{\odot}$}	
\newcommand{\cpd}{\,d$^{-1}$}	

\author{Amadeusz Miszuda}[UWR]
\author{Wojciech Szewczuk}[UWR]
\author{Jadwiga Daszy\'nska-Daszkiewicz}[UWR]
\affil[UWR]{Astronomical Institute, University of Wroc\l aw, ul. Kopernika 11, 51-662 Wroc\l aw, Poland}

\title{Simultaneous binary and asteroseismic modelling of the Kepler system KIC 10661783 with the $\delta$ Scuti component.}

\headtitle{Simultaneous binary and asteroseismic modelling of KIC 10661783}

\begin{document}

\maketitle

\begin{abstract}

		
		
		We present the results of analysis and modelling of the eclipsing binary system, KIC 10661783.
		The Fourier analysis of the \textit{Kepler} light curve, corrected for the binary effects, reveals 750 frequency peaks in both p and g-mode regions. Those with the highest amplitudes concentrate in the range of $20 - 30$\cpd. 
		
		To reproduce observed spectrum of frequencies we construct seismic models accounting for the mode instability. In order to obtain instabilities in the g-mode regime we modify the opacity tables data near the Z-bump.
		
		In order to reproduce system parameters, we construct evolutionary models including binary evolution.
		
		


\end{abstract}

\section{Introduction}

		KIC 10661783 is a close binary system of spectral type A5IV. It was studied for the first time by \cite{Pigulski2009}, who reported that its light curve exhibits eclipses of both components.

		The system was observed by the \textit{Kepler Space Telescope} for over four years. Its \textit{Kepler} light curve was analysed by \cite{Southworth2011}. The authors in short cadence (SC) Q2.3 and long cadence (LC) Q0-1 observations, found 68 pulsational frequencies. 
		According to them, all observed frequencies originate from the primary component.
		
		Later, \citet{Lehmann2013} using previously gathered 85 spectra of the system determined the mass-ratio of components. Its higher value than previously proposed by \cite{Southworth2011} suggested that the system is a post-mass transfer detached binary. Combining spectroscopic data and \textit{Kepler} photometric observations the authors determined the fundamental stellar parameters of KIC\,10661783 to be $M_A=2.100 \pm 0.028$ \mass, $R_A=2.575 \pm 0.015\,R_{\odot}$ for the primary and $M_B=0.1913 \pm 0.0025$ \mass, $R_B=1.124 \pm 0.019\,R_{\odot}$ for the secondary component.
		
\section{Observations}

\begin{figure}
	\centering
	\includegraphics[width=\textwidth,clip]{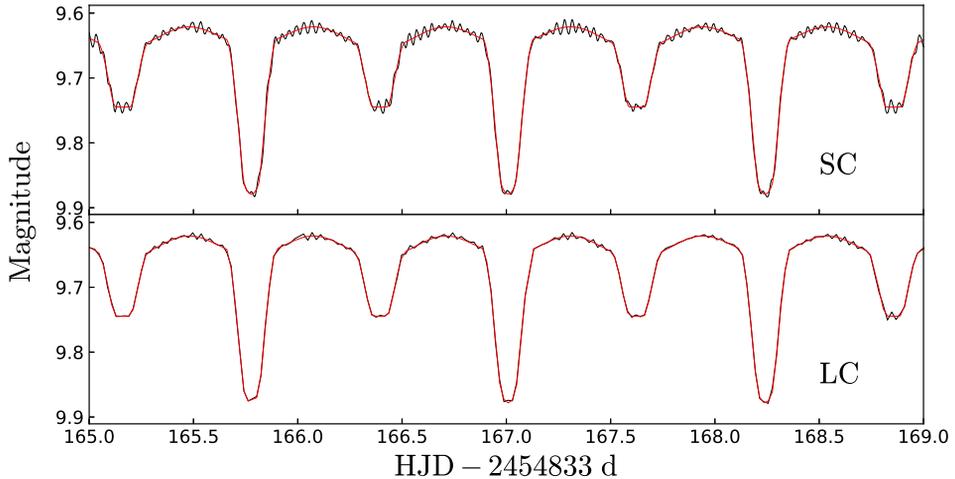}
	
	\caption{The comparison between \textit{Kepler} short cadence (top panel) and long cadence (bottom panel) light curve of KIC 10661783. In order to visualize both binary and pulsation variability only four days of observations are shown. The calculated \texttt{WD} model is marked with red lines.}
	\label{lc_comp}
\end{figure}

		KIC 10661783 was observed almost continuously for over four years by the \textit{Kepler Space Telescope} in two observational modes: short cadence and long cadence. 
		Long cadence (LC, Q0-Q17) data set for KIC 10661783 consists of over 58\,000 points spread 
		through 1500 days while short cadence (SC, Q2.3, Q6.1-Q8.3 and Q10.1-Q10.3) data set contains over 470\,000 observations spread 
		over 769 days period. The comparison of light curve from both, SC and LC observations can be seen in Fig.\,\ref{lc_comp}.

\section{Binary modelling}

		Given much longer time span of the observations than \cite{Southworth2011,Lehmann2013} we improved the orbital period of the system using all available LC observations. At first, it was done by the means of the \texttt{JKTEBOP} code \citep{Southworth2004} in order to obtain a good fit in a reasonably short time. Next, with the new orbital period value we modelled the system using Wilson-Devinney \citep[\texttt{WD}, see][]{Wilson1971,Wilson1979} code in detached mode to correct the light curve for the orbital variability. We found a new orbital period value to be $P=1.23136326 \pm 0.00000003$\,d. The fitted model can be seen in Fig.\,\ref{lc_comp}.


\section{Frequency analysis}

\begin{figure}
	\centering
	\includegraphics[width=\textwidth,clip]{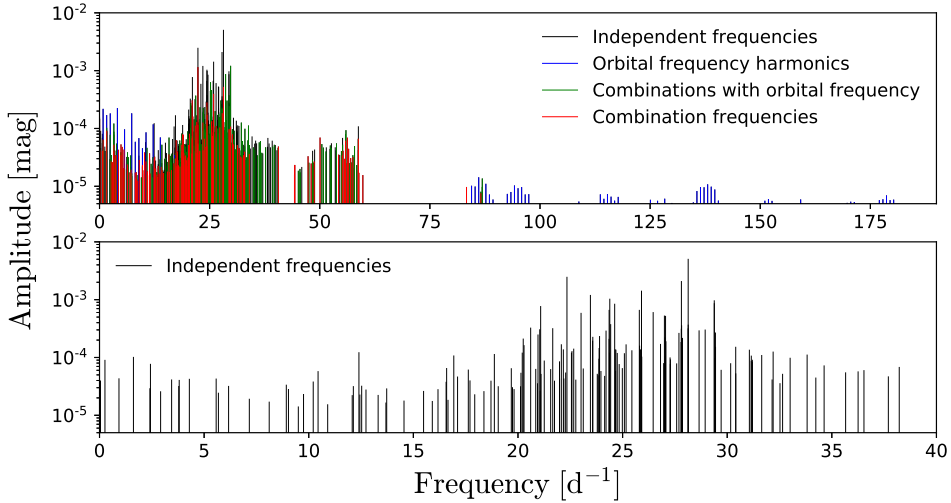}
	
	\caption{Results of the frequency analysis of the SC data after subtraction of the orbital model. In the top panel, we plotted all frequencies found from SC data. 
	Blue lines mark orbital frequency harmonics, green lines mark combinations with orbital frequency, red lines mark other combinations and black lines independent frequencies and those with no equivalent in LC data.
	The bottom panel shows only independent frequencies that have theirs equivalents in the LC data. Note that x-axis scales differ between panels.}
	\label{osc}
\end{figure}

		Fourier analysis of both, SC and LC data reveals a rich oscillation spectrum. Periodograms were calculated up to the Nyquist frequency for SC data ($\sim$680\cpd) and up to $200$\cpd\ for LC data with the iterative pre-whitening procedure.
		A careful analysis revealed 750 frequencies with the $S/N > 4.0$ for the SC data with the signal-to-noise ratio calculated in a 1\,d window centred on the given peak. 
		Most of the observed frequencies from SC data are located in the range $0 - 60$\cpd. Above $85$\cpd\ we observe only high combinations of orbital frequency (up to $223 \times f_{\rm orb}$). No periodicity was found above $200$\cpd. 
		Naturally, in the case of LC data there is a problem with aliasing due to the low Nyquist frequency value but this data gives better resolution and found frequencies can be compared with those found from SC analysis.
		Using a simple procedure of finding combination frequencies ($m \times f_i + n \times f_j$) and orbital harmonics ($N \times f_{\rm orb}$) we determined, that 167 amongst them seem to be independent with the accuracy of the Rayleigh resolution. 
		Frequencies which were found to be independent occupy both p- and g-mode regions (see Fig.\,\ref{osc}). 

		The low frequency pulsations in $\delta$ Scuti stars face a problem. The current standard models predict those frequencies to be stable. In order to fix that problem, opacity modifications near the Z-bump were proposed \citep{Pamyatnykh2004}. 
		
\begin{figure}[t]
	\centering
	\includegraphics[width=\textwidth,clip]{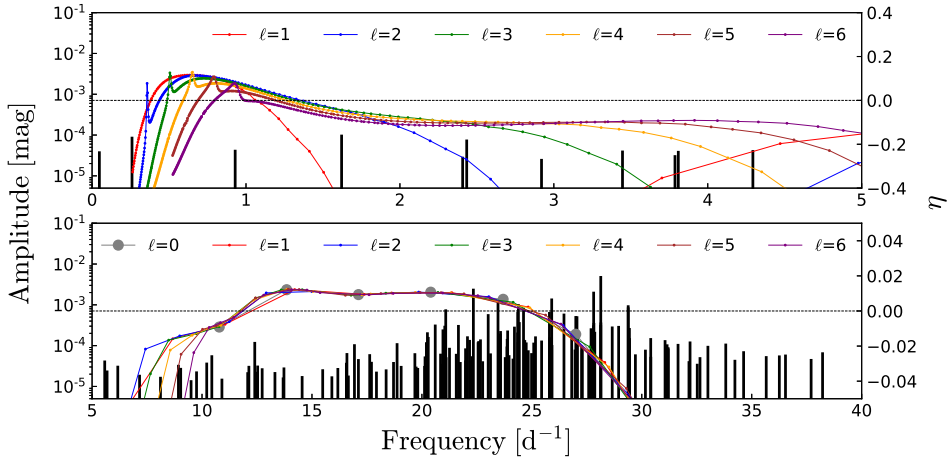} 
	
	\caption{Observed independent frequencies with the over-plotted theoretical pulsation models with the normalized instability parameter, $\eta$ for the representative model of KIC 10661783. The model was calculated using OPAL \citep{Iglesias1996} modified opacity tables for $\ell=0-6$ as an attempt to reproduce observed spectrum of frequencies (mainly in the g-mode domain). For clarity we plot only centroids.}
	\label{opacity}
\end{figure}

		In order to obtain unstable g-modes in the model of KIC 10661783 we modified OPAL \citep{Iglesias1996} opacity tables by increasing the opacity near the $\log T_{\rm eff}~=~5.06$ by 300\%. The results can be seen in Fig.\,\ref{opacity}, in which models for $\ell=0-6$ were plotted. 
		If we take into account rotational splitting we obtain unstable modes covering the whole g-mode region.
		

\section{Binary evolution}

		Algol-type binaries are not well described by the single star evolution models, since a lot of binary interactions are not taken into account.
		To model KIC 10661783 binary evolution with mass transfer we used the \texttt{MESA} code \citep{Paxton2011,Paxton2013,Paxton2015,Paxton2018,Paxton2019} with the \texttt{MESA-binary} module.
		
		Our evolutionary models including mass-transfer are able to reproduce masses of both components as well as the position of the primary on the H-R diagram. We failed to reproduce the position of the secondary and system's orbital period. Therefore we are not able yet to properly establish the evolutionary status and age of the KIC 10661783. More studies are required to unravel its status.

\section{Conclusions}

		In our work we present the analysis of all available \textit{Kepler} photometric data of KIC 10661783. Using this data we redetermined orbital elements. Based on the Fourier SC data analysis we found 750 frequencies in the range of $0 - 200$\cpd\ from which 167 seem to be independent. The highest-amplitude peaks concentrate in the range of $20 - 30$\cpd. In addition to frequencies typical for $\delta$ Sct pulsators we found low frequency signal that we interpret as high-radial order g-modes.

		We constructed seismic models accounting for the mode instability. In order to obtain g-mode excitation in theoretical models we modified the opacity tables data near the Z-bump. This helped to increase the value of instability parameter value, but the model is still far from satisfactory and further studies are needed.

		Finally, we constructed evolutionary models including binary evolution which reproduce masses of components and position of the primary on the H-R diagram but other observables are not well reproduced. In the future we plan to tune this model in order to reproduce all "binary" as well as pulsational observables.

\acknowledgements{
	This work was financially supported by the Polish National Science Centre grant 2018/29/B/ST9/02803.
	
	Calculations have been carried out using resources provided by Wroclaw Centre for
	Networking and Supercomputing (http://wcss.pl), grant no. 265.
	
	Funding for the Kepler mission is provided by the NASA Science Mission directorate.
	Some of the data presented in this paper were obtained from the Multi mission
	Archive at the Space Telescope Science Institute (MAST). STScI is operated by the
	Association of Universities for Research in Astronomy, Inc., under NASA contract.}

\bibliographystyle{ptapap}
\bibliography{miszuda}

\begin{thebibliography}{13}
\providecommand{\natexlab}[1]{#1}
\providecommand{\url}[1]{\texttt{#1}}
\providecommand{\urlprefix}{URL }
\providecommand{\eprint}[2][]{\url{#2}}

\bibitem[{{Iglesias} \& {Rogers}(1996)}]{Iglesias1996}
{Iglesias}, C.~A., {Rogers}, F.~J., \emph{\apj} \textbf{464}, 943 (1996)

\bibitem[{{Lehmann} et~al.(2013){Lehmann}, {Southworth}, {Tkachenko}, \&
  {Pavlovski}}]{Lehmann2013}
{Lehmann}, H., {Southworth}, J., {Tkachenko}, A., {Pavlovski}, K., \emph{\aap}
  \textbf{557}, A79 (2013)

\bibitem[{{Pamyatnykh} et~al.(2004){Pamyatnykh}, {Handler}, \&
  {Dziembowski}}]{Pamyatnykh2004}
{Pamyatnykh}, A.~A., {Handler}, G., {Dziembowski}, W.~A., \emph{\mnras}
  \textbf{350}, 3, 1022 (2004)

\bibitem[{{Paxton} et~al.(2011)}]{Paxton2011}
{Paxton}, B., et~al., \emph{\apjs} \textbf{192}, 1, 3 (2011)

\bibitem[{{Paxton} et~al.(2013)}]{Paxton2013}
{Paxton}, B., et~al., \emph{\apjs} \textbf{208}, 1, 4 (2013)

\bibitem[{{Paxton} et~al.(2015)}]{Paxton2015}
{Paxton}, B., et~al., \emph{\apjs} \textbf{220}, 1, 15 (2015)

\bibitem[{{Paxton} et~al.(2018)}]{Paxton2018}
{Paxton}, B., et~al., \emph{\apjs} \textbf{234}, 2, 34 (2018)

\bibitem[{{Paxton} et~al.(2019)}]{Paxton2019}
{Paxton}, B., et~al., \emph{\apjs} \textbf{243}, 1, 10 (2019)

\bibitem[{{Pigulski} et~al.(2009){Pigulski}, {Pojma{\'n}ski}, {Pilecki}, \&
  {Szczygie{\l}}}]{Pigulski2009}
{Pigulski}, A., {Pojma{\'n}ski}, G., {Pilecki}, B., {Szczygie{\l}}, D.~M.,
  \emph{\actaa} \textbf{59}, 1, 33 (2009)

\bibitem[{{Southworth} et~al.(2004){Southworth}, {Maxted}, \&
  {Smalley}}]{Southworth2004}
{Southworth}, J., {Maxted}, P.~F.~L., {Smalley}, B., \emph{\mnras}
  \textbf{351}, 4, 1277 (2004)

\bibitem[{{Southworth} et~al.(2011)}]{Southworth2011}
{Southworth}, J., et~al., \emph{\mnras} \textbf{414}, 3, 2413 (2011)

\bibitem[{{Wilson}(1979)}]{Wilson1979}
{Wilson}, R.~E., \emph{\apj} \textbf{234}, 1054 (1979)

\bibitem[{{Wilson} \& {Devinney}(1971)}]{Wilson1971}
{Wilson}, R.~E., {Devinney}, E.~J., \emph{\apj} \textbf{166}, 605 (1971)

\end{thebibliography}

\end{document}